\documentclass[preprint,12pt]{elsarticle}

\usepackage{amssymb}
\usepackage{amsmath}
\usepackage{booktabs}
\usepackage{threeparttable}

\newcounter{bla}

\journal{Computer Physics Communications}

\usepackage{tabularx}
\usepackage{graphicx}
\begin{document}

\begin{frontmatter}



\title{Effective bond-orbital model of III-nitride wurtzite structures based on modified interaction parameters of zinc-blende structures}


\author[a]{Fu-Chen Hsiao\corref{author}}
\author[b]{Ching-Tarng Liang}
\author[b]{Yia-Chung Chang}
\author[a]{John M. Dallesasse}

\cortext[author] {Corresponding author at: Department of Electrical and Computer Engineering, University of Illinois at Urbana-Champaign, Urbana, Illinois 61801. Tel.: +1 217 333 8416.\\\textit{E-mail address:} fhsiao3@illinois.edu}
\address[a]{Department of Electrical and Computer Engineering, University of Illinois at Urbana-Champaign, Urbana, Illinois 61801,USA}
\address[b]{Research Center for Applied Sciences, Academia Sinica, Taipei 11529, Taiwan}

\begin{abstract}

A simple theoretical method for deducing the effective bond-orbital model (EBOM) of III-nitride wurtzite (WZ) semiconductors is presented. In this model, the interaction parameters for zinc-blende (ZB) structures are used as an initial guess for WZ structure based on the two-center approximation. The electronic band structure of III-nitride WZ semiconductors can hence be produced by utilizing this set of parameters modified to include effects due to three-center integrals and fitting with first-principles calculations. Details of the semi-empirical fitting procedure for constructing the EBOM Hamiltonian for bulk III-nitride WZ semiconductors are presented. The electronic band structure of bulk AlN, GaN, and InN with WZ structure calculated by EBOM with modified interaction parameters are shown and compared to the results obtained from density functional (DFT) theory with meta-generalized gradient approximation (mGGA). The set of parameters are further optimized by using a genetic algorithm. In the end, electronic band structures and electron (hole) effective masses near the zone center calculated by the proposed model with best fitting parameters are analyzed and compared with the $\mathbf{k}\cdot\mathbf{p}$ model.

\end{abstract}


\end{frontmatter}


\section{Introduction}

III-nitride binary compounds, AlN, GaN, InN and their alloys, have been realized as remarkable materials in the fields of electronic and optoelectronic devices due to their unique electrical and optical properties, including large direct band gap \cite{AlNbg} and fairly wide range of emission frequency \cite{LargeR}. Such attributes make III-nitrides promising materials for high power devices and full-color display applications. Several novel optoelectronic devices have been reported based on III-nitride materials incorporated within nanostructures \cite{InNwell}--\cite{InNdot}.

As for the crystal structure, III-nitrides can crystallize in either WZ or ZB phase with the hexagonal WZ phase being thermodynamically more stable. It is possible to fabricate III-nitride semiconductors and corresponding nanostructures in either of the two crystal structures, depending upon growth conditions \cite{WZZBco}. In order to investigate the electronic and optical properties of III-nitride heterostructures and nanostructures, a model Hamiltonian capable of describing electronic band structures of constituent bulk materials is required. Although the density functional theory (DFT) \cite{DFT_1}--\cite{DFTning} can provide an accurate description of the dispersion relations in the full Brillouin zone (BZ), it has the well-known problem of underestimating the band gap \cite{DFTeg} and the immense computational complexity makes DFT an unsuitable approach for the modeling of large systems (with more than $10^3$ atoms per unit cell). The empirical tight-binding model (ETBM) \cite{TB1,TB2} requires much less computational effort than DFT, but a tedious fitting procedure is needed in order to determine a large number of empirical parameters. The $\mathbf{k}\cdot\mathbf{p}$ model \cite{KP1,KP1_1}, which makes use of the envelope function approximation, is the most widely adopted approach for modeling heterostructures due to its simplicity. It can incorporate multi-band effects near a band extremum and it has been used to study many problems with various modifications \cite{KP2}--\cite{KP4}. However, the inter-valley tunneling effects, which may be important for devices under high voltage operation, are difficult to handle due to the nature of $\mathbf{k}\cdot\mathbf{p}$ model. In addition, the ambiguity introduced by the ordering of the operators can sometimes lead to spurious solutions \cite{KPsp1,KPsp2} which causes instability of the model, especially in high-indium content devices. Based on the reasons above, the effective bond-orbital model (EBOM), which contains the virtues of ETBM and is comparable with the $\mathbf{k}\cdot\mathbf{p}$ model in terms of computational effort, has been proposed for both ZB and WZ structures \cite{EBOM1}--\cite{EBOM5}. This method has been shown to be computationally efficient, easy to implement, and particularly suitable for modeling device characteristics. In Ref. [24], second-neighbor interactions were used in order to fit the first conduction band and the top three valence bands over the entire BZ, but the resulting band structures do not have good accuracy. In this paper, we introduce an effective bond-orbital model (EBOM) for III-nitride WZ semiconductors with 17 nearest-neighbor interaction parameters, which are directly linked to those for their corresponding ZB semiconductors, plus 5 dominant second-neighbor interaction parameters involving $s$-like orbital. This procedure makes it easy to determine the interaction parameters, and the resulting band structures for the lowest conduction band and the top three valence bands are in fairly good agreement with DFT calculations with corrected band gaps.

\section{Effective bond-orbital model review}

In the EBOM, a minimum number of effective bond-orbitals are adopted to describe the most relevant portion of the band structure for bulk materials of which the nanostructure in the device is composed. For most of the semiconductors with WZ structures, two $s$-like anti-bonding orbitals and two sets of $p$-like bonding orbitals $(x, y, z)$ per unit cell would be sufficient to describe the band structure in the energy range of interest (i.e. near the band gap). The simple analytical expression \cite{EBOM6} of the EBOM Hamiltonian for the ZB structure allows for parameterization in terms of band energies at the high symmetry points. In this way, the band structures from EBOM are in good agreement with those obtained by a more rigorous model throughout the entire BZ while the electron (hole) effective masses remain fairly close to experimental values. With regard to the WZ structure, the direct parameterization of EBOM Hamiltonian targeting full BZ fitting is relatively difficult due to its lower crystal symmetry.

The general formulas in EBOM are briefly reviewed here in order to set the notation employed in this paper. Consider a collection of L\"{o}wdin functions $\vert\psi_{\alpha,i},\mathbf{R}_j\rangle$, denoting an $\alpha$-like bond-orbital centered on the lattice site $\mathbf{R}_j$ where index $i$ labels the plane perpendicular to the $c$-axis in WZ structure. According to Slater and Koster \cite{LCAO}, a set of Bloch sums, which serve as basis states for the tight-binding Hamiltonian, can be written as
\begin{equation}
\vert\psi_{\alpha,i},\mathbf{k}\rangle=\frac{1}{\sqrt{N}}\sum_{\mathbf{R}_j}e^{i\mathbf{k}\cdot\mathbf{R}_j}\vert\psi_{\alpha,i},\mathbf{R}_j\rangle
\label{eq_1},
\end{equation}
where $N$ denotes the number of lattice sites and $\mathbf{k}$ is the wavevector which lies in the first BZ.
The matrix elements of the tight-binding Hamiltonian can hence be written as 
\begin{equation}
H_{\alpha\alpha'}^{ii'}(\mathbf{k})\equiv\langle\psi_{\alpha,i},\mathbf{k}\vert H\vert\psi_{\alpha',i'},\mathbf{k}\rangle=\sum_{\mathbf{R}_j}e^{i\mathbf{k}\cdot\mathbf{R}_j}\langle\psi_{\alpha,i},0\vert H\vert\psi_{\alpha',i'},\mathbf{R}_j\rangle
\label{eq_2}.
\end{equation}
From Eq. (\ref{eq_2}), the definition of interaction parameter is given by 
\begin{equation}
E_{\alpha\alpha'}^{ii'}\equiv\langle\psi_{\alpha,i},0\vert H\vert\psi_{\alpha',i'},\mathbf{R}_j\rangle
\label{eq_3}.
\end{equation}
The interaction parameters are normally treated as adjustable parameters and determined by fitting with accurate energies at particular $\mathbf{k}$ values. If the concept of effective bond-orbitals is imposed on the set of L\"{o}wdin functions above, the tight-binding Hamiltonian reduces to the one adopted in EBOM. Notice that the information that comes from one of the sublattices has automatically been encoded in the effective bond-orbitals. As a result, instead of applying the full WZ structure, one merely considers a hexagonal close packed (HCP) structure in the EBOM Hamiltonian which significantly reduces the computational effort. 

\section{Interaction parameters for WZ structures based on modified parameters for ZB structures}

Now the goal turns to evaluating the matrix element of the EBOM Hamiltonian for the WZ structure. First, the interaction parameters of the EBOM for ZB structure are obtained by using formulas derived in Ref. [25] with the choice of fitting the conduction band energy at $X$ or $L$ point (whichever is lower in energy). In order to fit the lowest conduction band energy at $X$ and $L$ points simultaneously, an extra interaction parameter, $E_{sz}$, which describes the lack of inversion symmetry, is included in the EBOM for III-nitride semiconductors with a ZB structure. In these modified EBOM formulas, when considering III-nitride semiconductors, all the equations for calculating the interaction parameters are identical with those proposed in Ref. [25] (case B), while $E_{sz}$ is calculated by using the following constraint
\begin{equation}
E_{sz}=\sqrt{\frac{(E_c(L)-E_c(X)-4E_{ss})\chi}{48}}
\label{eq_1s}.
\end{equation}
Here, $\chi=E_{c}-12E_{ss}-E_v'+8(E_{xx}+E_{xy})+4E_{zz}$ while $E_c$ and $E_v'$ are the band edge positions for conduction and valence bands at zone center in the absence of the spin-orbit interaction. The values of $E_{g}$, $a$, $m_e$, and $\Delta_{so}$ used as input parameters are taken from Ref. [2] and Ref. [27]. The other input parameters, including heavy-hole band energies at $X$ and $L$ points, $E_{HH}(L)$ and $E_{HH}(X)$, the light-hole band energy at $X$ point, $E_{LH}(X)$, and the transition energies at $X$ and $L$ points, $E_{1}=(E_c(L)-E_{HH}(L))$ and $E_{2}=(E_c(X)-E_{HH}(X))$, are adopted from the values calculated by DFT with meta-generalized gradient approximation (mGGA) (i.e. using WIEN2K \cite{WIEN2K} with TB09 \cite{DFT_TB09} code and the experimental values of lattice constants).

\begin{table}[h]
\centering
\begin{threeparttable}
\caption{EBOM parameters for ZB III-nitrides used in this work. The symbol $E_{\alpha\alpha'}$ denotes the interaction between an $\alpha$-like bonding orbital at the origin and an $\alpha'$-like bonding locate at $(1,1,0)(a/2)$, where $a$ being the lattice constant in ZB structure. The remaining interaction parameters can be related to those in the table by $T_d$ point group. Energies are in eV.}
\begin{tabular}{lccc}
\hline
\hline
Parameter & \multicolumn{1}{c}{AlN} & \multicolumn{1}{c}{GaN} & \multicolumn{1}{c}{InN}\\
\hline
$E_{g}$ at $T=0$ K & 5.40\tnote{a} & 3.299$^{a}$ & 0.78$^{a}$ \\
$E_{g}$ at $T=300$ K & 5.34\tnote{b}  & 3.24$^{b}$  & 0.76$^{b}$  \\
$a$ (\AA) at $T=300$ K & 4.373\tnote{c} & 4.51$^{c}$ & 5.01$^{c}$\\
$m_{e}^{*}/m_{0}$ & 0.25$^{a}$ & 0.15$^{a}$ & 0.07$^{a}$ \\
$\Delta_{so} $ & 0.019$^{a}$ & 0.017$^{a}$ & 0.005$^{a}$ \\
$E_{1}$ & 9.2601 & 6.9500 & 4.9970 \\
$E_{2}$ & 6.5351 & 6.8870 & 6.0775 \\
$E_{HH}(L)$ & -0.4513  & -0.8933 & -0.7759 \\
$E_{HH}(X)$ & -1.6247  & -2.4974 & -2.0768 \\
$E_{LH}(X)$ & -4.4679  & -5.5309 & -4.5169 \\
$E_{ss}$ & 0.0271 & -0.0856 & -0.2028\\
$E_{sx}$ & 0.5248 & 0.4857 & 0.3051\\
$E_{sz}$ & 0.8850 & 0.6220 & 0.4183\\
$E_{xx}$ & 0.2792 & 0.3457 & 0.2823\\
$E_{zz}$ & -0.0758 & -0.0335 & -0.0227\\
$E_{xy}$ & 0.3699 & 0.4345 & 0.3479\\
\hline
\hline
\end{tabular}
\begin{tablenotes}
      \item[a]{Recommended value taken from Ref. [2].}
      \item[b]{Calculated value by solving the Varshni formula given the parameters from Ref. [2].}
      \item[c]{Experimental value taken from Ref. [27].}
    \end{tablenotes}
\label{TB_1s}
\end{threeparttable}
\end{table}

\begin{figure}[h]
		\centering\includegraphics[width=0.6\columnwidth]{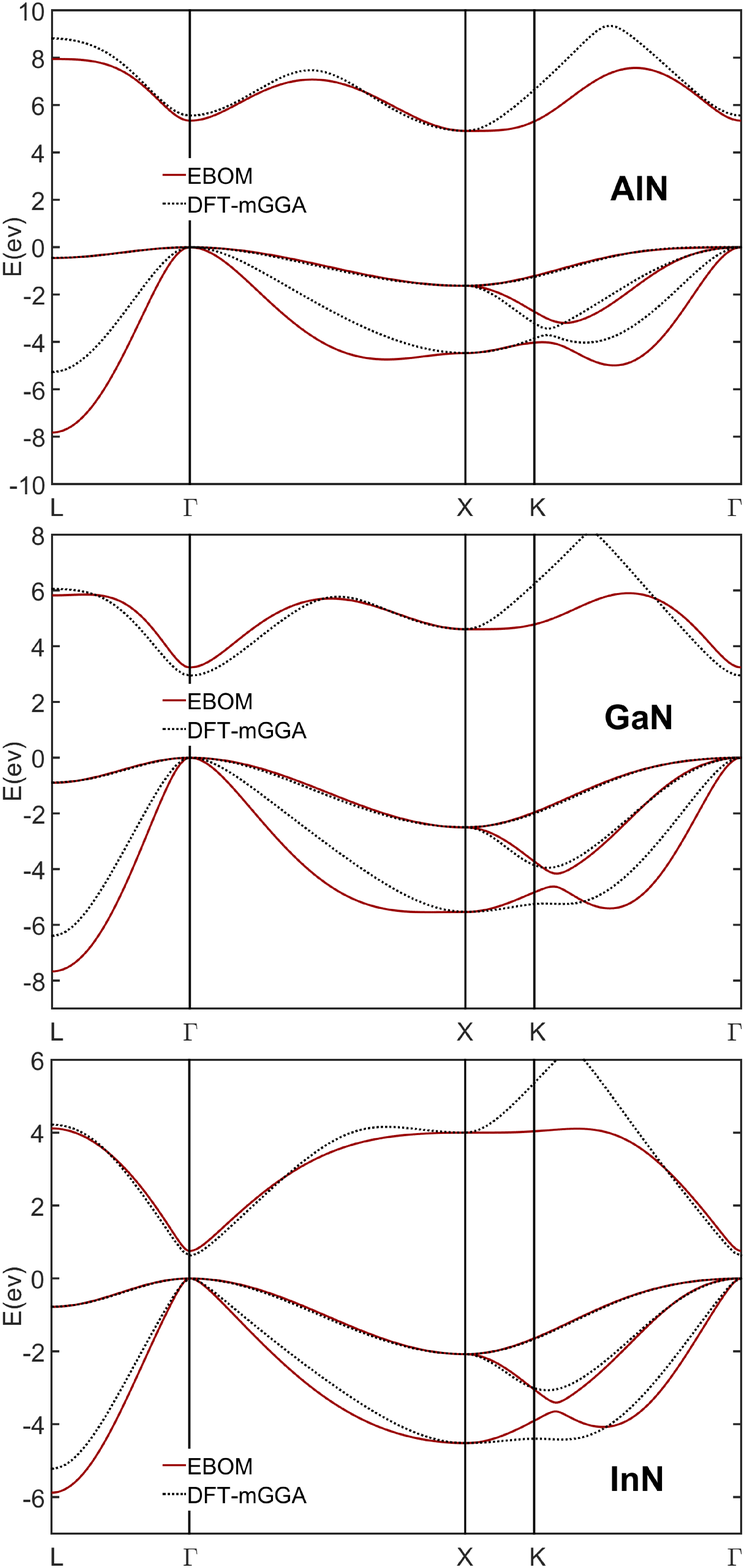}
	\caption{Band structures of AlN, GaN, and InN with ZB structure (from top to bottom) calculated by EBOM (solid curve) and DFT-mGGA (dotted curve).}
\label{fig_1s}
\end{figure} 

Following the modified formulas of EBOM with ZB structure described above, a set of interaction parameters between nearest-neighbor effective bond-orbitals for III-nitide ZB semiconductors at room temperature are obtained. All the input parameters and the calculated interaction parameters for AlN, GaN, and InN with ZB structure are listed in Table \ref{TB_1s}, where the temperature dependence of the bandgap at $\Gamma$ point is described by using the Varshni formula \cite{Varsh}. The resulting band structures at 300K for AlN, GaN, and InN with ZB structure derived from parameters in Table \ref{TB_1s} are shown in Fig. \ref{fig_1s}, while the comparison to those obtained by DFT-mGGA is presented as well.

For a given set of interaction parameters for ZB structures, the one for WZ structures can be calculated by utilizing the two-center approximation. In the two-center approximation, interaction integrals between two orbitals sitting on nearest neighbors as defined in Eq. (\ref{eq_3}) can be related to one another via a simple rotation (Slater transform \cite{LCAO}). As a consequence, all the interaction parameters associated with sites $0$ and any neighboring lattice site $\mathbf{R}_j$ can be expressed in terms of a set of parameters which are defined between sites $0$ and $\mathbf{R}_1=(1,1,0)a/2$ in ZB phase ($a$ being the lattice constant defined in ZB phase). Due to the difference of the stacking type between WZ and ZB structures (i.e. ABA stacking along $c$-axis in WZ structure while ABC stacking along [111]-direction in ZB structure), a constant scaling factor is multiplied with the initial interaction parameters obtained in the ZB structure to form the proper initial guess for the WZ structure. The scaling factors are empirically taken as $1.4$ for AlN, $1.3$ for GaN, and $1.2$ for InN in order to obtain the experimentally observed energy spacing between the first (3-fold) and second (2-fold) valence bands at the zone center for the WZ III-nitrides. Given the initial guess of the interaction parameters, the EBOM Hamiltonian with nearest-neighbor interaction for WZ structure is evaluated according to Eq. (\ref{eq_2}). The spin-orbit coupling effect is neglected in this paper, since its effect is very small ($<20$ meV \cite{LargeR}) for III-nitrides and the procedure of adding the spin-orbit effect is rather simple in EBOM \cite{EBOM6}.                   

The EBOM Hamiltonian for WZ structures under the two-center approximation possesses simple closed-form expression at high symmetry points. The analytical solutions for each high symmetry point are hence obtained by diagonalizing the Hamiltonian at corresponding wave vectors. By demanding the band energies at high symmetry points to be identical with the values calculated by DFT-mGGA, a set of equations are derived as follows:

\begin{subequations}
\label{eq_4}
\begin{equation}
E_{xx}^{12}=\frac{1}{4}(-\xi_{L}+\xi_{M}+\xi_{A})
\label{eq_4a},
\end{equation}
\begin{equation}
E_{yy}^{12}=\frac{1}{12}(3\xi_{L}-3\xi_{M}+\xi_{A})
\label{eq_4b},
\end{equation}
\begin{equation}
3E_{yy}^{11}+E_{xx}^{11}=\frac{1}{2}(\xi_{L}-\xi_{A})
\label{eq_4c},
\end{equation}
\begin{equation}
E_{xx}^{11}-3(E_{zz}^{11}+E_{zz}^{12})=\frac{1}{4}(2\Delta_{cr}-\xi_{L}-\xi_{A})
\label{eq_4d}.
\end{equation}
\end{subequations}

In Eqs. (\ref{eq_4a})--(\ref{eq_4d}), all the interaction parameters on the left-hand side are defined as in Eq. (\ref{eq_3}) with $\mathbf{R}_j=(a,0,0)$ for $i=i'=1$ and $\mathbf{R}_j=(0,a/\sqrt{3},c/2)$ for $i=1$ and $i'=2$, with $a$ and $c$ being the lattice constants defined in WZ structure. The symbol $\xi_{\gamma}$ denotes the band energy at the high symmetry point $\gamma$ in the BZ ($\gamma=L,M,A$) with respect to the valence band maximum (VBM) as calculated by DFT-mGGA, while $\Delta_{cr}$ is the crystal splitting energy. From Eq. (\ref{eq_4a}) and (\ref{eq_4b}), the exact solutions for $E_{xx}^{12}$ and $E_{yy}^{12}$ are obtained, which normally deviates from the initial guess by less than 20 meV. Eq. (\ref{eq_4c}) together with Eq. (\ref{eq_4d}) is insufficient to give unique solutions for $E_{xx}^{11}$, $E_{yy}^{11}$, $E_{zz}^{11}$, and $E_{zz}^{12}$. However, they can serve as constraints imposed during the fitting process in order to guarantee that the top valence band energies at all special symmetry points (except for the $K$ point) are identical with those calculated by DFT-mGGA. 

The actual procedures for deducing the remaining parameters are described as follows. Let $\eta=(3E_{yy,0}^{11}+E_{xx,0}^{11})-\frac{1}{2}(\xi_{L}-\xi_{A})$, where $E_{yy,0}^{11}$ and $E_{xx,0}^{11}$ are the values obtained via the relation based on two-center integral approximation. Since the top valence band energy at the $H$ point from the EBOM Hamiltonian under the two-center approximation is a function of $(E_{xx}^{11}+E_{yy}^{11})$, we then set $E_{yy}^{11}=E_{yy,0}^{11}-\eta/2$ and $E_{xx}^{11}=E_{xx,0}^{11}+\eta/2$ such that Eq. (\ref{eq_4c}) is satisfied, while avoiding influencing the top valence band energy at $H$ point. Next, $E_{zz}^{11}$ (for GaN and InN) or $E_{zz}^{12}$ (for AlN) are adjusted to satisfy Eq. (\ref{eq_4d}). The different selections are due to the ordering of the top three valence band energy at the $\Gamma$ point (i.e. the sign of the crystal splitting energy). The choice is also made based on the empirical fitting strategy which benefits the fitting at both $H$ and $M$ points, as well as the conduction band effective mass. Changing $E_{yy}^{11}$ and $E_{zz}^{11}$ both can break the $T_d$ symmetry but not the $C_{3v}$ symmetry which is compatible with the WZ structure. Finally, minor adjustment of $E_{yz}^{12}$ is required  to fit the top valence band at the $H$ point. Note that the top valence band energy at the $K$ point is at least $1.95 $ eV below the VBM, which plays a less important role in determining the device characteristics. Therefore, the $K$ point is not considered in the current fitting procedure. The above fitting procedure will cause a shift of the VBM. This can be remedied by readjusting the on-site energies to bring the band energy at VBM back to zero.

In the present model, the effect of three-center integrals can be introduced by imposing symmetry relations between interaction parameters according to the $C_{3v}$ point group for the WZ structure, instead of the rotational group for the two-center approximation or the $T_d$ point group for the ZB structure. As a result, the three-center effect can be incorporated in the EBOM Hamiltonian by adding extra interaction parameters that break certain symmetry relations for $T_d$ group among original parameters. For now, the three-center effects are turned on in the current model only if it is necessary to fit more subtle features in the band structures. In this paper, the objective values of crystal splitting energies are assigned to be $-0.169$ eV, $0.01$ eV, and $0.04$ eV for AlN, GaN, and InN, respectively. The band gap energies at $T=300$ K are taken as $6.16$ eV, $3.44$ eV, and $0.76$ eV for AlN, GaN, and InN, respectively. All the parameters above are directly adopted or deduced from the values recommended by Vurgaftman \cite{LargeR}. Notice that for GaN, the three-center related parameter $E_{xz}^{11}$, which corresponds to the Luttinger-like parameter $A_{7}$, is included in order to fit the anticrossing behavior between the light-hole and split-off band (due to crystal-splitting) in the $\mathbf{k}\cdot\mathbf{p}$ model. On the other hand, modifying $E_{zz}^{11}$ and $E_{yy}^{11}$ is the only adjustment which corresponds to three-center effect in the EBOM calculation of InN and AlN.

\begin{figure}[h]
		\centering\includegraphics[width=0.6\columnwidth]{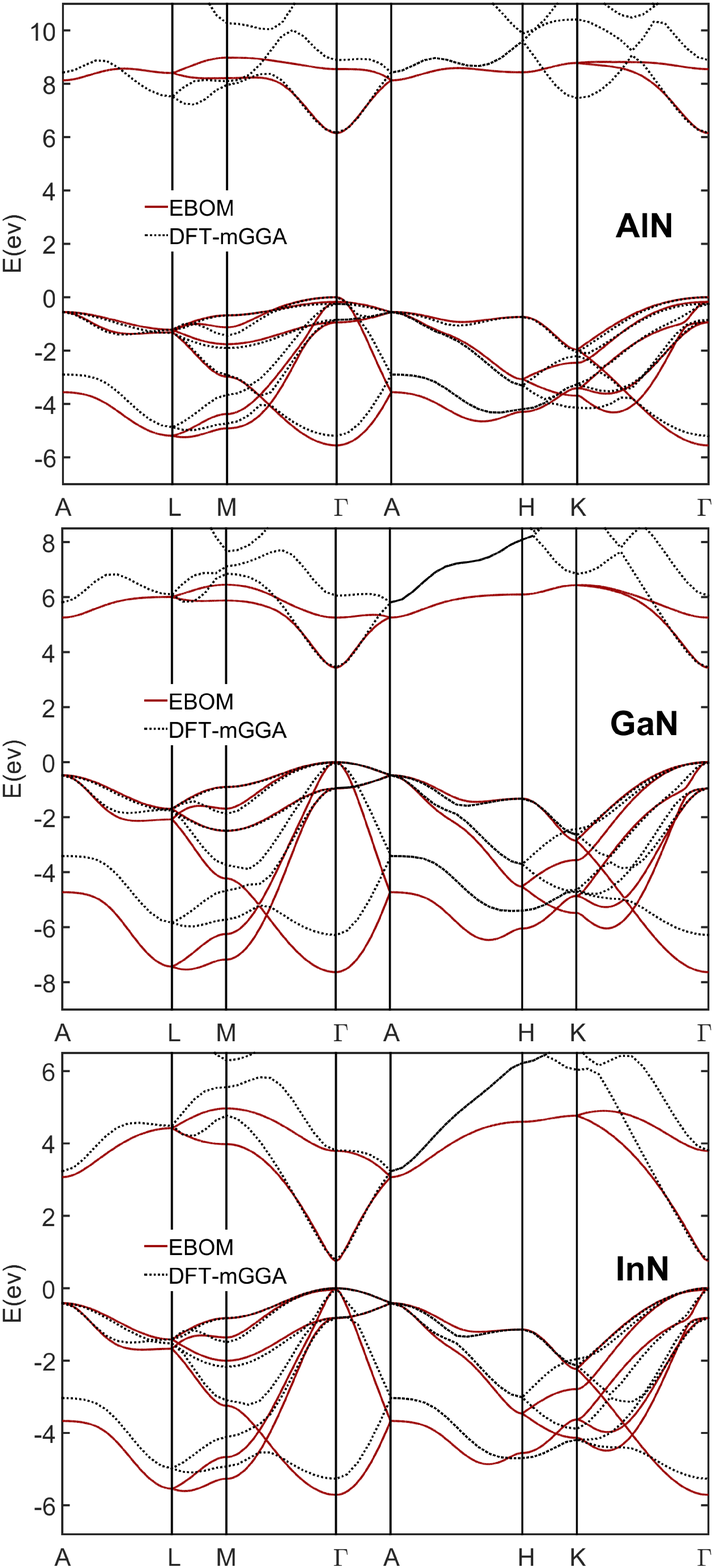}
	\caption{Band structures of AlN, GaN, and InN with WZ structure (from top to bottom) calculated by EBOM with modified interaction parameters (solid curve) and DFT-mGGA (dotted curve).}
\label{fig_1}
\end{figure}

\begin{figure}[h]
		\centering\includegraphics[width=0.6\columnwidth]{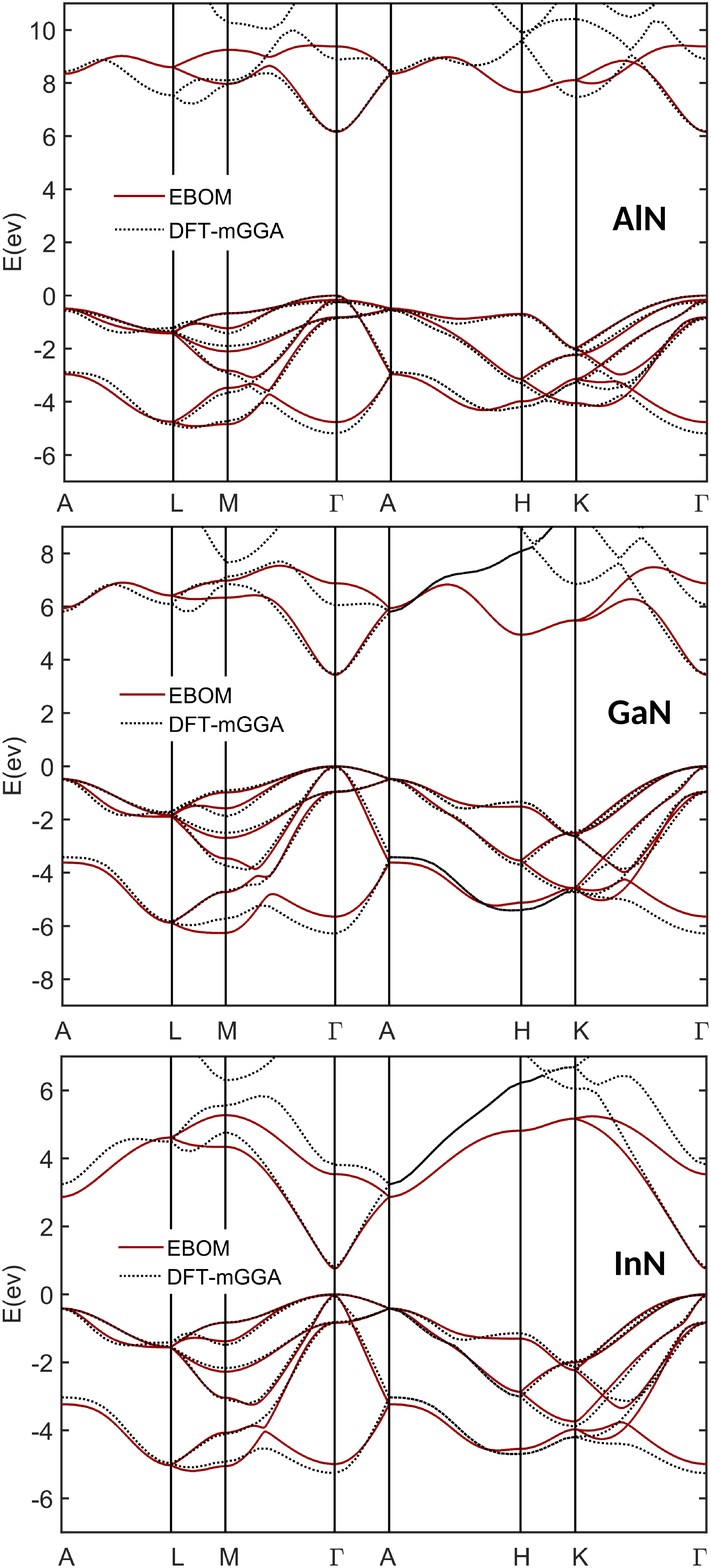}
	\caption{Band structures of AlN, GaN, and InN with WZ structure (from top to bottom) calculated by EBOM with best fitting interaction parameters (solid curve) and DFT-mGGA (dotted curve).}
\label{fig_3}
\end{figure}

\begin{table}[h]
\centering
\caption{EBOM parameters for WZ III-nitrides used in this work (without optimization). All the material parameters are adopted from Ref. [2]. The bandgap energy at $T=300$ K are calculated by using Varshni formula \cite{Varsh} given the parameters recommended by Vurgaftman \cite{LargeR}. The non-vanishing interaction parameters not given in the table can be related to the ones in the table by the two-center approximation. Energies are in eV.}
\begin{tabular}{lccc}
\hline
\hline
Parameter & \multicolumn{1}{c}{AlN} & \multicolumn{1}{c}{GaN} & \multicolumn{1}{c}{InN}\\
\hline
$a$ (\AA) at $T=300$ K & 3.112 & 3.189 & 3.545 \\
$c$ (\AA) at $T=300$ K & 4.982 & 5.185 & 5.703 \\
$E_{g}$ at $T=0$ K & 6.25 & 3.51 & 0.78 \\
$E_{g}$ at $T=300$ K & 6.16  & 3.44  & 0.76 \\
$\Delta_{cr}$ & -0.169 & 0.010 & 0.040 \\
$E_{ss}^{11}$ & -0.1202 & -0.1942 & -0.2581\\
$E_{sx}^{11}$ & 0.5773 & 0.4941 & 0.4061\\
$E_{xx}^{11}$ & 0.6723 & 1.0430 & 0.7659\\
$E_{yy}^{11}$ & -0.1138 & -0.1442 & -0.0883\\
$E_{zz}^{11}$ & -0.0906 & -0.1077 & -0.0714\\
$E_{xz}^{11}$ & 0.0000 & 0.0140 & 0.0000\\
$E_{ss}^{12}$ & -0.1992 & -0.1512 & -0.2531\\
$E_{sy}^{12}$ & 0.4775 & 0.4540 & 0.2768\\
$E_{sz}^{12}$ & 0.4864 & 0.4773 & 0.3626\\
$E_{xx}^{12}$ & -0.0387 & -0.0796 & -0.0433\\
$E_{yy}^{12}$ & 0.1672 & 0.2398 & 0.1810\\
$E_{zz}^{12}$ & 0.4623 & 0.6355 & 0.4724\\
$E_{yz}^{12}$ & 0.3988 & 0.5199 & 0.3753\\
\hline
\hline
\end{tabular}
\label{TB_1}
\end{table}

\begin{table}[h]
\centering
\caption{Best fitting EBOM parameters for WZ III-nitrides used in this work. All the material parameters are adopted from Ref. [2]. The bandgap energy at $T=300$ K are calculated by using Varshni formula \cite{Varsh} given the parameters recommended by Vurgaftman \cite{LargeR}. The non-vanishing interaction parameters not given in the table can be related to the ones in the table by $C_{3v}$ point group symmetry. Energies are in eV.}
\begin{tabular}{lccc}
\hline
\hline
Parameter & \multicolumn{1}{c}{AlN} & \multicolumn{1}{c}{GaN} & \multicolumn{1}{c}{InN}\\
\hline
$a$ (\AA) at $T=300$ K & 3.112 & 3.189 & 3.545 \\
$c$ (\AA) at $T=300$ K & 4.982 & 5.185 & 5.703 \\
$E_{g}$ at $T=0$ K & 6.25 & 3.51 & 0.78 \\
$E_{g}$ at $T=300$ K & 6.16  & 3.44  & 0.76 \\
$\Delta_{cr}$ & -0.169 & 0.010 & 0.040 \\
$E_{ss}^{11}$ & 0.0105 & 0.0238 & -0.3022\\
$E_{sx}^{11}$ & 0.4858 & 0.3504 & 0.3614\\
$E_{xx}^{11}$ & 0.5836 & 0.7623 & 0.6583\\
$E_{yy}^{11}$ & -0.0462 & -0.0281 & -0.0304\\
$E_{zz}^{11}$ & -0.0418 & -0.0235 & -0.0263\\
$E_{sy}^{11}$ & -0.1753 & -0.1919 & -0.0540\\
$E_{sz}^{11}$ &  0.0753 & 0.0260 & -0.0444\\
$E_{xy}^{11}$ & -0.0442 & 0.0472 & 0.0729\\
$E_{xz}^{11}$ & -0.0567 & -0.0142 & 0.1003\\
$E_{yz}^{11}$ & 0.0069 & 0.1481 & 0.0737\\

$E_{ss}^{12}$ & -0.0989 & -0.2260 & -0.2321\\
$E_{sy}^{12}$ & 0.4094 & 0.4964 & 0.3497\\
$E_{sz}^{12}$ & 0.4860 & 0.4204 & 0.2923\\
$E_{xx}^{12}$ & -0.0969 & -0.0930 & -0.0748\\
$E_{yy}^{12}$ & 0.2057 & 0.2537 & 0.2150\\
$E_{zz}^{12}$ & 0.3988 & 0.4704 & 0.4708\\
$E_{yz}^{12}$ & 0.3794 & 0.3912 & 0.3443\\

$E_{ss}^{(0,\sqrt{3}a,0)}$                       & -0.0258 & -0.1730 & -0.0831\\
$E_{ss}^{(a,\frac{\sqrt{3}a}{3},\frac{c}{2})}$   & -0.1708 & -0.0610 & -0.0071\\
$E_{ss}^{(\frac{3a}{2},\frac{\sqrt{3}a}{2},0)}$  & -0.1458 & -0.1573 & 0.0637\\
$E_{sy}^{(0,\frac{-2\sqrt{3}a}{3},\frac{c}{2})}$ & 0.1143 & 0.0366 & 0.0639\\
$E_{sz}^{(0,\frac{-2\sqrt{3}a}{3},\frac{c}{2})}$ & -0.0686 & 0.0179 & 0.0438\\
\hline
\hline
\end{tabular}
\label{TB_8}
\end{table}

\section{Interaction parameters for WZ structures optimized by genetic algorithm}
The set of interaction parameters obtained in the previous section now can be used as a set of good initial guesses for numerical fitting. Adding extra parameters by considering full $C_{3v}$ point group symmetry in the EBOM Hamiltonian, the interaction parameters are optimized by conducting the genetic algorithm \cite{GA} to fit the results from DFT-mGGA. In this case, all the three-center related parameters are available during the fitting procedure. In addition to nearest-neighbor interaction, since the less localized property of $s$-like orbital used to model the conduction band, second-neighbor interaction parameters between two $s$-like orbitals, $E_{ss}^{(x,y,z)}$, as well as one $s$-like and one $p$-orbital sitting on different plane in z-axis, $E_{s\alpha}^{(x,y,z)}$, $\alpha=x,y,z$, are included to improve the conduction band structure. Here, $(x,y,z)$ denotes the positions of orbitals corresponding to second-neighbor atoms. In this paper, $(0,\sqrt{3}a,0)$, $(3a/2,\sqrt{3}a/2,0)$, $(a,\sqrt{3}a/3,c/2)$, and $(0,-2\sqrt{3}a/3,c/2)$ are chosen to construct second-neighbor interaction parameters for the EBOM Hamiltonian. The relations between interaction parameters which obey the $C_{3v}$ point group symmetry are presented by a matrix in orbital basis as shown in Eq. (\ref{eq_5a}) and Eq. (\ref{eq_5b}), which illustrate the relation between the interaction parameters on the same/different plane perpendicular to the $c$-axis, respectively. Due to the glide plane reflection symmetry in WZ structures, the interaction parameters with the site of one orbital are identical while the other differs only by the sign on the z-axis, parameters can be simply related by adding a minus sign on either one. As for $E_{zz}$, the value remains the same in two sets of parameters since the double effects are compensated by each other. The same relations are applied to the second-neighbor interaction parameters.

During the optimization procedure via genetic algorithm, the goal is to fit the overall electronic band structure with those calculated by DFT-mGGA, and the conduction band effective masses with the values obtained from either reliable experimental measurements or DFT-mGGA. As recommended by Vurgaftman \cite{LargeR}, the electron effective masses for AlN in the $k_z$ and in-plane directions are $m^{\parallel}=0.31m_0$ and $m^{\perp}=0.30m_0$, while $m^{\parallel}=0.2m_0$ and $m^{\perp}=0.2m_0$ for GaN. For lack of reliable experimental studies, in the case of AlN, the objective values mentioned above have been assigned an equal weighting number with the ones deduced from DFT-mGGA, $m^{\parallel}=0.35m_0$ and $m^{\perp}=0.37m_0$, in the fitting process. Due to the absence of better information, the conduction band effective masses calculated by DFT-mGGA, $m^{\parallel}=0.08m_0$ and $m^{\perp}=0.09m_0$, are selected as objective values in the optimization fitting of InN.  
\begin{subequations}
\begin{equation}
\begin{bmatrix}
    E_{ss}^{}  & E_{sx}^{}  & E_{sy}^{}  & E_{sz}^{} \\
    -E_{sx}^{} & E_{xx}^{}  & E_{xy}^{}  & E_{xz}^{} \\
    E_{sy}^{}  & -E_{xy}^{} & E_{yy}^{}  & E_{yz}^{} \\
    E_{sz}^{}  & -E_{xz}^{} & -E_{yz}^{} & E_{zz}^{} \\
\end{bmatrix}
\label{eq_5a}
\end{equation}

\begin{equation}
\begin{bmatrix}
     E_{ss}^{}  & 0          & E_{sy}^{}  & E_{sz}^{} \\
     0          & E_{xx}^{}  & 0          & 0 \\
    -E_{sy}^{}  & 0          & E_{yy}^{}  & E_{yz}^{} \\
    -E_{sz}^{}  & 0          & E_{yz}^{}  & E_{zz}^{} \\
\end{bmatrix}
\label{eq_5b}
\end{equation}
\end{subequations}

\section{Results and discussion}
In Fig. \ref{fig_1}, the full band structure for bulk AlN, GaN, and InN with WZ structure calculated by EBOM with modified interaction parameters, which are deduced in Section 3, are shown and compared with those from DFT-mGGA. The material parameters as well as the modified interaction parameters used in this work are listed in Table \ref{TB_1}. As seen in Fig. \ref{fig_1}, the overall band structures obtained by the present model with modified interaction parameters before optimization are very close to those in DFT-mGGA for the energy range from $1.5$ eV below the VBM to 1.5 eV above the CBM (conduction band minimum). Furthermore, all energies of the top valence band at $\Gamma, L, M, A$ and $H$ are pinned at desired values. Fig. \ref{fig_3} shows the electronic band structure with the best fitting parameters optimized by the genetic algorithm, where the former interaction parameters are used as initial guesses. All the best fitting parameters used in this work are given in Table \ref{TB_8}. The overall band structure has been significantly improved to achieve fairly high similarity with the results from DFT-mGGA.     
 
\begin{figure}[t]
		\centering\includegraphics[width=0.8\columnwidth]{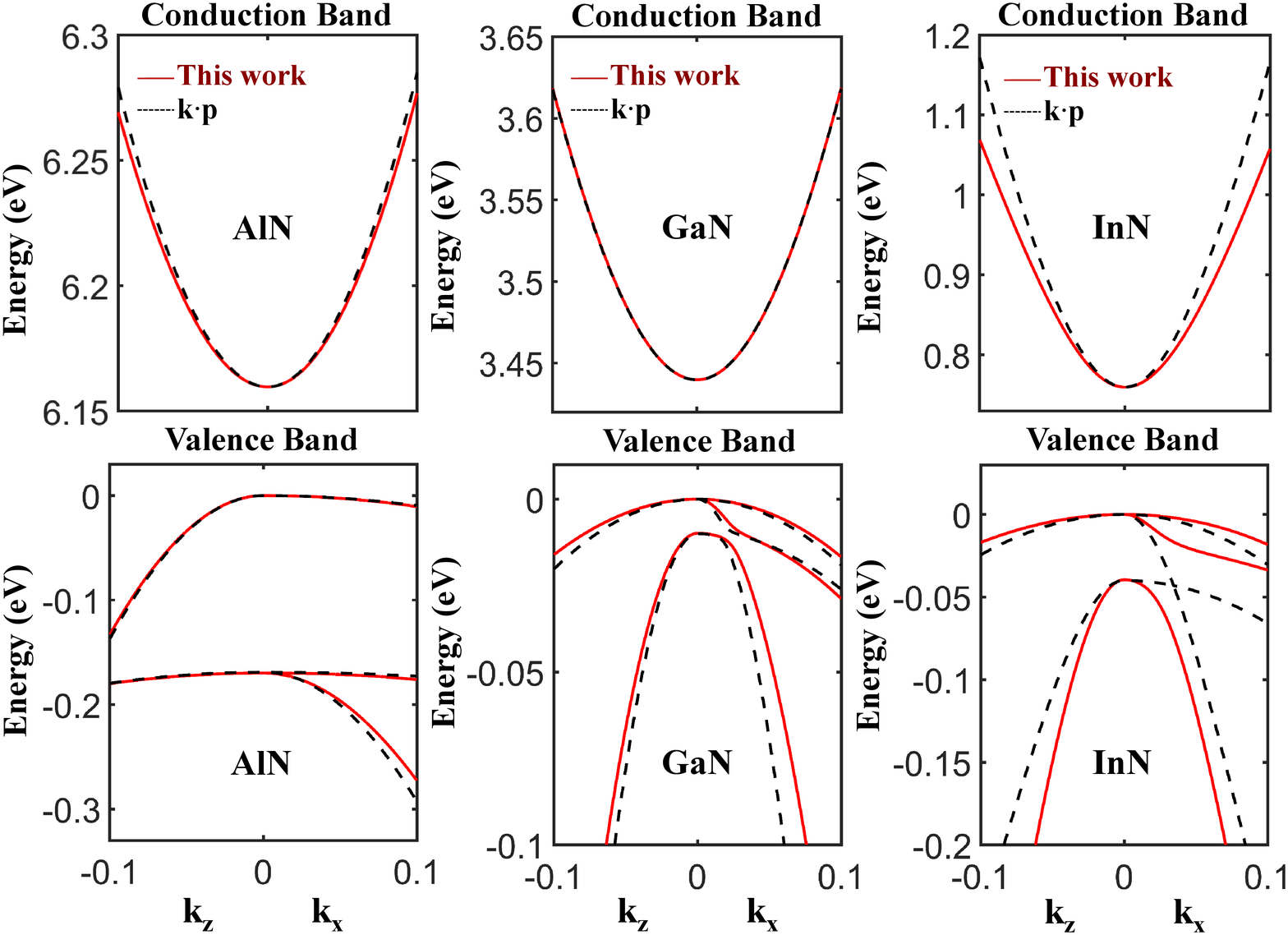}
	\caption{Band structures of III-nitrides with WZ structure calculated by EBOM with best fitting interaction parameters  (solid curve) and $\mathbf{k}\cdot\mathbf{p}$ model (dashed curve).}
	\label{fig_2}
\end{figure}

Fig. \ref{fig_2} shows a comparison between the bulk band structures for AlN, GaN, and InN with the WZ structure near $\Gamma$ point obtained by the EBOM (with best fitting interaction parameters) and the $\mathbf{k}\cdot\mathbf{p}$ model. As expected, the band structures obtained from the EBOM with best fitting parameters and the $\mathbf{k}\cdot\mathbf{p}$ model are nearly identical for small $\mathbf{k}$ values. For large $\mathbf{k}$ values, the current model gives more satisfactory band structures than the $\mathbf{k}\cdot\mathbf{p}$ model. The noticeable discrepancy in the valence band for InN is due to the neglecting of the Luttinger-like parameter $A_{7}$ in the $\mathbf{k}\cdot\mathbf{p}$ model. Although the recommended values for the electron effective mass of InN ($m^{\parallel}=0.07m_0$ and $m^{\perp}=0.07m_0$) are less than the ones predicted in DFT-mGGA, considering the quality of the InN sample, error of carrier concentration measurement, and the strain effect in the film experiment, the results obtained from DFT-mGGA are assigned to be the objective values during fitting in this work.    

The electron and hole effective masses at the zone center derived from EBOM with best fitting parameters are given in Table \ref{TB_2}. The electron effective masses for III-nitrides with WZ structure fit those in $\mathbf{k}\cdot\mathbf{p}$ model (for AlN, GaN) or DFT-mGGA (for InN), while the hole effective masses are within the range proposed by other theoretical models \cite{EM_1}--\cite{EM_5}. Both conduction and valence band effective masses have been further optimized toward the targeted values after the optimization fitting. Notice that further experimental studies are needed to determine the Luttinger parameters of InN. As a consequence, it is believed that the first principle calculation provides a more reliable band structure for now. In the present model with best fitting parameters, less than $7\%$ deviation from the DFT-mGGA results have been achieved for InN.   

\begin{table}[h]
\begin{threeparttable}
\caption{Electron and hole effective masses obtained in EBOM with best fitting interaction parameters, DFT-mGGA calculation, and $\mathbf{k}\cdot\mathbf{p}$ model by curve fitting. Units are in free electron mass.}
\begin{tabular}{cccccccccc}
\hline
\hline
& Ref. & $m^{\parallel}_{\text{e}}$ & $m^{\parallel}_{\text{hh}}$ & $m^{\parallel}_{\text{lh}}$ & $m^{\parallel}_{\text{ch}}$ & $m^{\perp}_{\text{e}}$ & $m^{\perp}_{\text{hh}}$ & $m^{\perp}_{\text{lh}}$ & $m^{\perp}_{\text{ch}}$ \\
\hline
AlN\\
 & This work & 0.34 & 3.76 & 3.76 & 0.28 & 0.32 & 7.69 & 0.36 & 3.71\\
 & DFT-mGGA & 0.35 & 3.67 & 3.67 & 0.32 & 0.37 & 11.7 & 0.41 & 4.43\\
 & $\mathbf{k}\cdot\mathbf{p}$\tnote{a} & 0.31 & 3.57 & 3.57 & 0.27 & 0.30 & 10.0 & 0.30 & 4.00\\
GaN\\
 & This work & 0.20 & 2.35 & 2.35 & 0.17 & 0.20 & 2.38 & 0.23 & 1.09\\
 & DFT-mGGA & 0.23  & 2.24 & 2.24 & 0.21 & 0.26 & 2.46 & 0.28 & 1.08\\
 & $\mathbf{k}\cdot\mathbf{p}$\tnote{a} & 0.20 & 1.89 & 1.89 & 0.14 & 0.20 & 2.00 & 0.15 & 1.11\\
InN\\
 & This work & 0.08 & 2.23 & 2.23 & 0.08 & 0.09 & 2.22 & 0.15 & 0.20\\
 & DFT-mGGA & 0.08 & 2.20 & 2.20 & 0.08 & 0.09 & 2.10 & 0.16 & 0.19\\
 & $\mathbf{k}\cdot\mathbf{p}$\tnote{a} & 0.07 & 1.56 & 1.56 & 0.12 & 0.07 & 1.25 & 0.10 & 1.47\\
\hline
\hline
\end{tabular}
\label{TB_2}
\begin{tablenotes}
      \item[a]{$\mathbf{k}\cdot\mathbf{p}$ calculation given the band parameters under room temperature in Vurgaftman \cite{LargeR}.}
    \end{tablenotes}
    \end{threeparttable}
\end{table}

\section{Conclusions}
In conclusion, a theoretical model for calculating the electronic band structure of III-nitride WZ semiconductors based on the EBOM with modified interaction parameters is presented. The same set of fitting parameters can be used for the parametrization of the EBOM Hamiltonian for both ZB and WZ structures given the relation provided by the two-center approximation. The interaction parameters are modified by fitting with the band structure obtained in DFT-mGGA at the high symmetry points. The three-center effect is partially incorporated for the cases with a WZ structure in order to fit more sophisticated features in the bands structure. The set of parameters hence is utilized as initial guesses for conducting numerical fitting procedures by using the genetic algorithm. The current model has been applied to calculate the band structures of III-nitrides with a WZ structure over the entire BZ and the results are in good agreement with those obtained from DFT-mGGA in the energy range suitable for device characterization. It is also shown that the proposed model can fit the desired electron effective masses, while the hole effective masses are within the range of values provided by other groups. 

\section{Acknowledgments}
This work is supported by the National Science Foundation (MRSEC program), under Grant DMR-1120923, 
the Ministry of Science and Technology, Taiwan, under Grant MOST 106-2112-M-001-022.


\end{document}